\documentclass[12pt]{article}
\usepackage{latexsym}

\let\ssection=\section
\renewcommand{\section}{\setcounter{equation}{0}\ssection}

\newcommand\mathC{\mkern1mu\raise2.2pt\hbox{$\scriptscriptstyle|$}
        {\mkern-7mu\rm C}} 
\newcommand{\mathR}{{\rm I\! R}}         

\newcommand\bi{\begin{itemize}}
\newcommand\ei{\end{itemize}}
\newcommand\be{\begin{equation}}
\newcommand\ee{\end{equation}}

\newcommand{\dd}{\ensuremath{{\delta}}}
\newcommand{\al}{\ensuremath{{\alpha}}}

\begin{document}
\begin{titlepage}

\begin{center}
{\large\bf On the Persistence of Particles}
\end{center}

\vspace{0.2 truecm}

\begin{center}
        J.~Butterfield\footnote{email: jb56@cus.cam.ac.uk;
            jeremy.butterfield@all-souls.oxford.ac.uk}\\[10pt] All Souls 
College\\ Oxford OX1 4AL
\end{center}

\begin{center}
       Dedicated to the memory of Jim Cushing, an amazing mind and a wonderful 
man.\\
  Forthcoming in {\em Foundations of Physics}: 21 January 2004 
\end{center}

\vspace{0.2 truecm}

 \begin{abstract}\noindent   
This paper is about the metaphysical debate  whether objects persist over time 
by the selfsame object existing at different times (nowadays called  `{\em 
endurance}' by metaphysicians), or by different temporal parts, or stages, 
existing at different times (called `{\em perdurance}'). I aim to illuminate 
the debate by using some elementary kinematics and real analysis: resources 
which metaphysicians have, surprisingly, not availed themselves of. There are 
two main results, which are of interest to both endurantists and 
perdurantists.

 (1): I describe a precise formal equivalence between the way that the two 
metaphysical positions  represent the motion of the objects of classical 
mechanics (both point-particles and continua).

 (2): I make precise, and prove a result about, the idea that the persistence 
of objects moving in a void is to be analysed in terms of tracking the 
continuous  curves in spacetime that connect points occupied by matter. The 
result is entirely elementary: it is a corollary of the Heine-Borel theorem. 
\end{abstract}

\end{titlepage}
 \setcounter{page}{1}
 
\section{Introduction}\label{IntrCush}
In this paper I will address a debate in metaphysics, using some resources of 
elementary mathematics (kinematics and analysis): resources which 
metaphysicians have, surprisingly, not availed themselves of. The metaphysical 
debate is about the persistence  of objects over time: does an object persist 
over time by the selfsame object existing at different times (nowadays called  
`{\em endurance}' by metaphysicians), or by different temporal parts, or 
stages, existing at different times (called `{\em perdurance}')? 

I will describe the two rival positions ({\em endurantism} and {\em 
perdurantism}) in more detail in Section \ref{sec;perseintrCush}. Then I use 
some elementary mathematics to give two results about how these two positions 
describe the objects of classical mechanics: results which should be of 
interest to both positions. Here I use `object' to include both:\\
\indent (i) point-particles, which we think of as moving in a void (so that a 
system composed of finitely many point-particles is finite-dimensional); and\\
\indent (ii) classical continua, i.e. bodies whose composing matter entirely 
fills their volume, so that the body is strictly speaking 
infinite-dimensional; though it may be small and-or rigid enough to be treated 
as finite-dimensional. (Indeed, it may be small and rigid enough to be treated 
as a point-particle: as in the usual Newtonian mechanical  treatment of 
planetary motion!)

\indent The first result (in Section \ref{intertransCush}) is a precise formal 
equivalence between the way that endurantism and perdurantism represent the 
motion of  objects: both point-particles and continua, in either a classical 
or relativistic spacetime. (But I will agree that because the equivalence is 
formal, it is liable to be broken by various philosophical considerations.)

\indent The second result (in Section \ref{ppsCush}) make precise the idea 
that the persistence of  objects moving in a void is to be analysed in terms 
of tracking the continuous  curves in spacetime that connect points occupied 
by matter. (The result is entirely elementary: it is a corollary of the 
Heine-Borel theorem.) By and large, it is perdurantists, not endurantists, who 
discuss this idea of analysing persistence in terms of tracking matter; since 
for them, persistence is not identity, so that they need to tell us what they 
take it to be. (This endeavour  is  called `defining the genidentity relation 
between temporal parts', as well as `analyzing persistence'.)  But I maintain 
that this result (like the first) is not only of interest for perdurantists: 
for endurantists, it makes precise the idea of keeping track of enduring 
objects as they move through space.

\indent On the other hand, this second result is limited in a way that the 
first is not: as follows. One of the main arguments in the metaphysical 
literature on persistence  is an argument  against perdurantism, that turns on 
the contrast between point-particles and continua.  The  argument is based on 
two ideas:\\
\indent (i) {\em Homogeneous}:  In a continuum (i.e. continuous body) that is 
utterly homogeneous throughout a time-interval containing two times $t_0,t_1$, 
a spatial part at the time $t_0$ is equally  qualitatively similar to any 
spatial part congruent to itself (i.e. of the same size and shape) at the 
later time $t_1$. (The properties of the continuum can change over time, but 
must not vary over space; e.g. the continuum could cool down, but must at each 
time have the same temperature everywhere.)\\
\indent (ii) {\em Follow}: The perdurantist will presumably try to analyze 
persistence (define genidentity) in terms of following timelike curves of 
maximum qualitative similarity.\\
\indent The strategy of {\em Follow} seems to work well   when applied to 
point-particles moving  in a void with a continuous spacetime trajectory 
(worldline): for starting at a point-particle at $t_0$, there is a unique 
timelike curve of qualitative  similarity passing through it. Similarly for 
point-particles moving, not in a void, but in a continuous fluid with suitably 
different properties (a different ``colour'', or made of different ``stuff''). 
That is rough speaking: but it is widely accepted---and I will also accept it.  
But {\em Homogeneous} implies that  {\em Follow}'s strategy stumbles when 
applied to a homogeneous continuum. There are altogether too many spatial 
parts at $t_1$ that are tied-first-equal as regards qualitative similarity to 
the given spatial part at $t_0$: any congruent spatial part will do. In other 
words:   the curves of qualitative similarity run ``every which way''.\\
\indent This problem is  made vivid by urging that the perdurantist  cannot 
distinguish two  cases that, the argument alleges, must be distinguished: for 
example, a perfectly circular and rigid disc of homogeneous matter that is 
stationary, and a duplicate disc (congruent, rigid, homogeneous, and made of 
the same material) that is rotating.  Hence the argument  is nowadays often 
called the `rotating discs argument'; (recent discussions, including 
references, include: Hawley 2001, p. 72-90, Sider 2001 p. 224-236).  

I believe that the perdurantist can rebut this argument. But I argue this 
elsewhere (2004, 2004a), and will not take up the issue. Here, I need to say 
only that this paper's second result is limited in the sense that it does not 
address the argument. Nor does it concern the idea that underlies the 
argument, of qualitative similarity relations among the spatial parts of a 
homogeneous body. In other words, the second result applies primarily to 
point-particles moving in a void (or in a fluid of different ``stuff''). It is 
applied to extended bodies by ignoring their inner constitution: i.e. in 
effect by treating them as small and rigid enough to be modelled as a 
point-particle. But as the above example of a planet shows, this need not be a 
drastic limitation.

I turn to summarizing how these results bear on the metaphysical dispute 
between endurantism and perdurantism. First: I  do not claim that these 
results resolve the  dispute, even for the objects of classical mechanics; nor 
that the sort of technical resources I use could somehow be exploited to do 
so. Indeed, even when the metaphysical dispute is considered only in 
connection with the objects of classical mechanics, the considerations on the 
two sides are so various and inter-related as to resist formalization. So  we 
cannot be sure that  the dispute can be formulated in a precise way acceptable 
to both sides: let alone expect it to  have  a definitive, maybe formal or 
technical,  resolution. (Details in Section \ref{sec;perseintrCush}; though   
I daresay  almost any philosophical dispute similarly resists a definitive, 
and in particular  a technical, resolution! Cf. e.g.  Kripke (1976, Section 
11(b), p. 407-416).)

 But I do claim that my results shed some light on the dispute. I think that 
(i) the  equivalence contained in the first result, and (ii) the fact that the 
second result's  idea of tracking can be endorsed by the endurantist as well 
as the perdurantist, suggests  that---as regards describing the motion of the 
objects of classical mechanics---the honours are about even between 
endurantism and perdurantism. `Honours about even' is a vague and ecumenical 
conclusion. But I think it is a worthwhile one---especially since, as I argue 
elsewhere, it can be supported by a rebuttal of the rotating discs argument. 
And because of that rebuttal, I believe that overall,  as regards classical 
mechanics, the cases for endurantism and perdurantism are about equally 
strong. In any case, setting aside this paper's results: I suggest that  the 
sort of  resources I use here are a promising armoury for attacking various 
somewhat technical questions about persistence that the philosophical  
literature seems not to have addressed.

Finally, a point about my discussion's being restricted to classical 
mechanics: and its focussing (in the second result) on classical  
point-particles, or objects small and rigid enough to be modelled as 
point-particles. This restriction implies, I am afraid, that (unlike most 
papers in this memorial issue) I will not discuss any of the mysteries of 
quantum theory: not even in that lucid and even unmysterious form, the 
pilot-wave theory, that Cushing championed so persuasively (1994; 1998 
Chapters 23, 24). But I hope that this mixture of physics and philosophy, and 
more specifically, this scrutiny of point-particles, honours the memory of a 
man who had such mastery of both these disciplines---and whose work gave 
point-particles such a good press!

\section{Endurantism Vs. Perdurantism}\label{sec;perseintrCush}
I begin by introducing the metaphysical debate about persistence.
I adopt the following terminology, which is now widespread. `Persistence' is 
the neutral word for the undeniable fact that objects are not instantaneous: 
objects, or at least most objects,  exist for a while. The debate is over how 
to understand persistence.  The two main positions can be roughly stated as 
follows:---\\
\indent (i): {\em Endurantism}: The endurantist holds that  persistence is a 
matter of the selfsame object being present at two times---as is often said 
for emphasis, `wholly present at two times'. This is called `endurance'. At 
first sight, this view seems  close to ``common sense'', and on that account 
plausible: it seems that the selfsame rock exists at noon and 12.05.\\
\indent (ii): {\em Perdurantism}: The perdurantist holds that at each time, 
only a stage or phase of the object is present: persistence is a matter of 
there being a sequence of  ``suitably related'' stages. At first sight, this 
view clashes with ``common sense'', and is to that extent implausible: a rock 
seems not to have stages. So perdurantists urge that their view has advantages 
over endurantism that are worth the cost of revision; some perdurantists say 
the advantages dictate the revision, others just that they favour it. I will 
not need to list, let alone assess, these alleged advantages: they concern 
both general principles, and the solutions to various puzzle cases.

 Another way to put the contrast between endurantism and perdurantism is in 
terms of spatial and temporal parts of objects. Both the endurantist and 
perdurantist accept that objects have spatial parts; e.g. the arm of the sofa. 
The perdurantist urges that they likewise have temporal parts, viz.  stages.  
Again, such temporal parts seem to clash with common sense; and accordingly 
perdurantists admit that their view is revisionary.\footnote{But even 
perdurantists should accept that spatial and temporal parts are individuated 
differently; for details, cf. Butterfield (1985, pp. 35-37).} 

This terminology of `persistence', `endurance' and `perdurance' is due to 
Johnston, and became widespread through  the influence of its adoption by 
Lewis (1986, pp. 202-204). Another widespread terminology is to call 
endurantism `three-dimensionalism' and perdurantism `four-dimensionalism': a 
terminology which is adopted by Sider, whose fine monograph (2001) surveys the 
debate and defends (what he calls!) four-dimensionalism. Sider (2001, p.3) 
helpfully lists many advocates on both sides. Among the recent advocates he 
lists are the following:\\
\indent (i) for endurantism: Haslanger (1994), Johnston (1987),  Mellor (1998) 
and Rea (1998); Rea (1998)  is  useful since he replies {\em seriatim} to 
various arguments against endurantism;\\
\indent (ii) for perdurantism: Balashov (1999), Hawley (2001), Lewis (1986, 
pp. 202-204; 1988, 1999), and Sider himself.

I admit that my formulations of endurantism and perdurantism above are  
sketchy; (and consequently, so is my claim that endurantism is closer to 
common sense). In fact, the advocates listed give various formulations; and 
both sides of the debate have found problems in their opponents' 
formulations---even to the extent of saying they do not understand them! Thus 
some endurantists have complained that temporal parts are problematic, or even 
unintelligible; and some perdurantists have found the formulation of endurance 
problematic. (For example, Sider discusses both allegations, and endorses the 
second; 2001, pp.53-62 and 63-68 respectively.)

But in this paper, I shall not  need to be very precise about the formulation 
of either position.  I can make do with the following adaptation of Sider's 
position. In short: I will follow his philosophical methodology, and his 
formulation of perdurantism; but I will deny his allegation that endurantism 
is problematic. 

Sider takes it that the perdurantist believes in the existence of a kind of 
object, viz. temporal parts, that the endurantist (i) understands but (ii) 
denies to have any instances. (He defends the presupposition here that the two 
parties disagree about a matter of fact, albeit one that is hard to know, 
against a ``no-conflict view'' of the type suggested by Carnap's linguistic 
frameworks.) Sider then formulates  perdurantism in terms intelligible by the 
endurantist, as follows; (2001, pp.55-62). The endurantist accepts the notions 
of existence at a time, and one object being a part of another at a time. The 
claim that there are instantaneous temporal parts of any (spatiotemporal as 
against abstract) object is then:
\begin{quote} 
for any object $o$, for any time $t$ at which $o$ exists, there is an object 
that: (i) exists only at $t$; (ii) is part of $o$ at $t$; (iii) overlaps any 
object that is part of $o$ at $t$.
\end{quote}  
(Clause (iii) secures that the temporal part encompasses $o$'s entire spatial 
extent at $t$.) Similarly for non-instantaneous temporal parts, corresponding 
to any interval $[t_1,t_2]$ throughout which $o$ exists. The claim is then:
\begin{quote} 
for any object $o$, for any interval $[t_1,t_2]$ throughout which $o$ exists, 
there is an object that, for any time $t \in [t_1,t_2]$: (i) exists at $t$; 
(ii) is part of $o$ at $t$; (iii) overlaps any object that is part of $o$ at 
$t$.
\end{quote}
(One could make stipulations about what to say about intervals $[t_1,t_2]$ for 
which $o$ does not exist throughout the interval.)

\indent So endurantism denies these claims. Since they are universally 
quantified, such a denial could be very weak, claiming only that some object 
does not have all the temporal parts (instantaneous or extended) claimed by 
the perdurantist. Of course, endurantists usually make a much stronger denial. 
They say that the objects of ordinary ontology---J.L. Austin's `medium-sized 
dry goods' like chairs, organisms like people, and ``wet goods'' like lakes or 
clouds---have {\em no} temporal parts. This is how I shall understand 
endurantism.

 To which I add four comments, of which the third and fourth are most relevant 
for what follows:\\
\indent  (i): Sider himself argues that formulating  endurantism is 
problematic. (He does this by stating and rebutting possible meanings for the 
endurantist's catchphrase that objects are ``wholly present at a time''; 2001, 
pp. 63-68.) But I think endurantism is adequately formulated as just the  
suitably strong denial, of the perdurantist's claims above. 

\indent  (ii): My formulation of the debate (and adjustments one might make in 
the light of (i)) brings out that the debate is metaphysical, not linguistic 
or epistemic. From this perspective, some arguments in the literature, in 
particular objections to perdurantism, are misdirected. (Cf. Sider (2001, 
p.208-212) for some robust replies along these lines. More positively, this 
means that the perdurantist can adopt a number of different views about the 
relation of temporal parts to temporal language and its semantics; which Sider 
also discusses.)

\indent  (iii): There is of course a compromise {\em mixed view}, that some 
objects such as meals and explosions---objects one might call `events'---have 
temporal parts. Though I think this view has much to recommend it (2004a), I 
will not explicitly discuss it below: for it will be obvious how my 
discussion, in particular my two results, would apply to this view.\\
\indent But I should  stress that the mixed view has a significant background 
role in my discussion. For I will assume that endurantists and perdurantists 
alike can talk of the spacetime manifold, and spacetime regions of various 
types like timelike curves (worldlines) and spacelike slices. I defend this 
assumption elsewhere (2004). Here it must suffice to say that many on both 
sides of the current endurantism-perdurantism debate  are ``scientific 
realists'', and even substantivalists about spacetime. They believe that 
successful scientific theories like relativity theory, literally construed, 
are approximately true; and even that spacetime points  are {\em bona fide} 
objects bearing the properties and relations represented by mathematical 
structures like metrics and connection.  Perdurantists who believe this 
typically take spacetime regions also to be perduring spatiotemporal objects, 
viz. mereological fusions of their points, rather than being  ``abstract'' 
sets of points; (where `abstract' means at least `spatiotemporally 
non-located, so that the question of persistence does not arise'). But since 
temporally extended  spatiotemporal regions surely do not endure, endurantists 
who are substantivalists must either (a) take regions to be abstract sets of 
points, or (b) adopt the mixed view, and then say that spatiotemporal regions 
are among the events.       

\indent  (iv): Elsewhere (2004a), I argue that the perdurantist can perfectly 
well advocate only extended i.e. non-instantaneous temporal parts; and that 
there is good reason for them to do so---i.e. to deny the existence of 
strictly instantaneous parts. For one thing, a perdurantist who advocates only 
extended  temporal parts has a complete reply to the rotating discs argument 
(2004). In (2004a), I also diagnose perdurantists' traditional acceptance of 
instantaneous parts as due to an erroneous doctrine I call `{\em 
pointillisme}'. In this paper, I will not need to be precise about {\em 
pointillisme}. I only need two main ideas; as follows.\\
\indent \indent First: roughly speaking, {\em pointillisme} is the claim that 
the history of the world is fully described  by all the intrinsic properties 
of all the spacetime  points and-or all the intrinsic properties at all the 
various times of point-sized bits of matter.\\
\indent \indent Second: here, `intrinsic'  means `both spatially and 
temporally intrinsic'; so that (a) attributing such a property carries no 
implications about matters at other places, or at other times, and (b) a {\em 
pointilliste} perdurantist  will  seek to analyse persistence as a matter of 
some suitable relations between instantaneous stages.\footnote{Two ancillary 
remarks. (1) I will not need to be precise about the intrinsic-extrinsic 
distinction among properties; which is fortunate, since  how best to 
understand it is controversial. (2) A  warning about my jargon: What I here 
call `{\em pointillisme}' is called in my (2004a) `{\em pointillisme} as 
regards spacetime'---to distinguish it from another doctrine, which is here 
irrelevant.} 

\section{An Equivalence}\label{intertransCush}
In this Section, my main aim is to present my equivalence between 
endurantism's and perdurantism's representations of the motion of the objects 
of classical mechanics (whether point-particles  or spatially 
extended).\footnote{One can think of the equivalence as making precise an 
accusation sometimes made by people (in my experience, especially physicists 
and mathematicians) when they are first  told about the distinction between 
endurance and perdurance---that it sounds spurious, a  difference only in 
words.}

 This equivalence  is ``only formal''---it ignores philosophical issues. And I 
agree that, as so often, the formal nature of the equivalence makes it liable 
to be broken by philosophical considerations. I shall not list, let alone 
survey,  all these considerations. (I discuss some of them elsewhere (2004a). 
And anyone familiar with the endurantism-perdurantism debate, e.g. as surveyed 
by Hawley (2001) and Sider (2001), will be able to list yet more.)  But I   
note that one  such consideration, viz. criteria of identity, is the topic of 
Section \ref{ppsCush}. And in this Section, the main point about breaking the 
equivalence will be that  under certain assumptions (including  the {\em 
pointilliste} assumption of using instantaneous temporal parts), the 
equivalence can fail  {\em   formally} for continua, i.e. continuous bodies 
(Section \ref{35:troublepism}). Besides, {\em aficionados} of the rotating 
discs argument will  recognize  this failure as the formal fact underlying 
that argument. Hence my  arguments in (2004, 2004a) that once {\em 
pointillisme} is rejected, the rotating discs argument fails, and perdurantism 
is tenable for the continuous bodies of classical mechanics.

 I will present the equivalence  in an informal way. I will not state more 
exactly than I did in Section \ref{sec;perseintrCush} what endurantism and 
perdurantism claim; nor what notions (like `... is a part of ... at time $t$') 
they each find intelligible. Nor will I need to say what exactly is intended 
by each of them ``recovering'' in their own terms, the perduring/enduring 
object advocated by the other side. This informality will have the advantage 
of clarity. But more important: it will also, I  think, strengthen the 
equivalence, by making it equally applicable to various precise formulations 
of endurantism and perdurantism that one might adopt. But I will not try to 
prove this. Though I agree that it would be a good project to do so, e.g. for 
the formulations discussed by Sider (2001, pp. 53-68), it is not necessary for 
this paper. 

\subsection{The idea: fusing worldlines}\label{31:fusingworldlines}
The equivalence depends on the idea that a spatially unextended i.e. 
point-sized enduring object can be mathematically represented by a worldline, 
i.e. a curve in spacetime giving its spatiotemporal location at any time at 
which it exists. Here a curve is defined as a function $q$ from time $t \in 
\mathR$ to spatiotemporal locations $q(t)$; so for a point-particle, $q(t) \in 
{\cal M}$, with $\cal M$ the manifold representing spacetime. (I shall shortly 
generalize the discussion to spatially extended objects.)

 On the other hand, a perduring spatially unextended object with its various 
stages can be mathematically represented by a collection of ``shorter'' 
worldlines, one for each stage.  That is, we represent the location of each 
stage by a function defined on the corresponding time-interval, mapping times 
to the stage's location at that time. 

The idea of the equivalence is now obvious. Any function $q$ defined on a 
domain ${\rm dom}(q)$ is equivalent  to (i.e. defines and is defined by) any 
collection of functions $q_{\al}$, $\al$ the index, on a collection of subsets 
of dom($q$) that (i)  cover dom($q$), i.e. $\cup_{\al} {\rm dom}(q_{\al}) = 
{\rm dom}(q)$, and (ii) mesh in the sense that if dom($q_{\al_1}$) and 
dom($q_{\al_2}$) overlap, dom($q_{\al_1}$) $\cap$ dom($q_{\al_2}$) $\neq 
\emptyset$, then $q_{\al_1}$ and $q_{\al_2}$ agree on the overlap:
\be
\forall t \in {\rm dom}(q_{\al_1}) \cap {\rm dom}(q_{\al_2}) \neq \emptyset \; 
, \;\;\; q_{\al_1}(t) = q_{\al_2}(t) \; .
\label{eq;qsmesh}
\ee

To make the point as simply as possible, let us consider an enduring 
point-particle, labelled $i$, that is eternal in that it exists at all $t \in 
\mathR$; and let us locate it in Euclidean space $\mathR^3$ rather than 
spacetime. So we have a  function $q_i:\mathR \rightarrow \mathR^3$. As to 
perdurantism, we will consider only stages corresponding to closed intervals 
of time $[a,b] \subset \mathR$. It will be clear that nothing in the sequel 
depends on dom($q_i$) being $\mathR$; I could instead fix a temporal interval 
large enough to include all objects and stages of objects to be considered. 
Nor will anything depend on using closed intervals, rather than say open ones. 
Nor will anything depend on using {\em all} closed intervals, rather than a 
family that cover the lifetime of the particle.

The function $q_i$ immediately defines suitable mathematical representatives 
of the perdurantist's stages, viz. the restrictions of $q_i$ to subsets of its 
domain; in particular, restrictions to closed intervals $[a,b] \subset 
\mathR$: $q_i \mid_{[a,b]}: t \in [a,b] \mapsto q_i(t) \in \mathR^3$. 
Conversely,  on the perdurantist conception of the point-particle, we 
represent the location of each stage by a function defined on the 
corresponding time-interval, mapping times to the stage's location at that 
time. So we  have, for each closed interval $[a,b] \subset \mathR$, a function  
$q_{[a,b]}: t \in [a,b]  \mapsto q_{[a,b]}(t) \in \mathR^3$; where these 
functions are now not given as restrictions of a function $q_i$. But suppose 
we are given such a collection of functions which agree with each other in the 
obvious way expected for a point-particle, viz. that if $t \in [a,b] \cap 
[c,d]$, then $q_{[a,b]}(t) = q_{[c,d]}(t)$. That is, the functions mesh in the 
sense of eq. \ref{eq;qsmesh}. Then there is a unique function $q_i:\mathR 
\rightarrow \mathR^3$ whose restrictions $q_i \mid_{[a,b]}$ to intervals 
$[a,b]$ are the given functions $q_{[a,b]}$.  This function $q_i$ is a 
suitable mathematical representative of the (location of) the endurantist's 
enduring object. (I have merely labelled the endurantist's function, but not 
the perdurantist's functions, with $i$.) This result suggests that the 
endurantist and perdurantist can each ``reconstruct'' what the other says 
about the persistence of an object.

\subsection{Generalizations}\label{32:generalztns}
This equivalence extends readily to much more general situations. It is 
obvious that it extends in the ways mentioned above: to a point-particle that 
is not eternal, and to stages not specified by a closed interval of times. 

Furthermore, for a point-particle, the perdurantist can work with {\em 
instantaneous} stages. (But beware: we will see in Section 
\ref{35:troublepism} that this extension to instantaneous stages will not work 
for continua, rather than point-particles---which will spell trouble for the 
{\em pointillisme} introduced at the end of Section \ref{sec;perseintrCush}.) 
For one way to cover the domain ${\rm dom}(q)$ of a function $q$ is with the 
singleton sets $\{t\}, t \in {\rm dom}(q)$. That is: suppose the endurantist 
writes for a point-particle labelled $i$ that is, say, eternal, a single 
function: $q_i: t \in \mathR \mapsto q_i(t) \in \mathR^3$. Then the 
perdurantist can write an uncountable family of functions $q_t$, one for each 
$t \in \mathR$, with domain just $\{t\}$, sending $t$ to the point-particle's 
location then: $q_t(t) \mapsto q_i(t) \in \mathR^3$. This uncountable family 
of functions trivially defines the endurantist's function $q_i$. (The meshing 
requirement eq. \ref{eq;qsmesh} is now vacuous; no two domains overlap.)   

The equivalence also obviously extends to spatiotemporal location, rather than 
spatial location. The simplest such extension conceives spacetime as just 
$\mathR \times \mathR^3$, and defines a new $q'_i(t) := (t,q_i(t))$; but the 
generalization using any spacetime manifold $\cal M$ as codomain of the 
function $q_i$ is immediate. In particular, the equivalence obviously extends 
to relativity. Indeed nothing in the discussion above requires the curve 
$q_i:\mathR \rightarrow {\cal M}$ to be non-spacelike (i.e. having a tangent 
vector that at all points on the curve is on or in the light-cone). 

It also extends to an object that is spatially extended. This requires that 
$q_i$ takes as values subsets, not points, of the spacetime  $\cal M$ (or of 
space, if we consider spatial not spatiotemporal location). But with $q_i$'s 
values thus adjusted, the equivalence holds good. Agreed, if we are 
considering spatiotemporal location, so that $q_i(t) \subset {\cal M}$, we 
will want $q_i(t)$ to be a spacelike set of points so as to represent an 
instantaneous location of the object; (`spacelike' can be defined for 
non-relativistic spacetimes, on analogy to the relativistic  definition). We 
will probably also want no two $q_i(t), q_i(t')$, for $t \neq t'$, to overlap. 
But such requirements do not affect the equivalence. 

Besides, these extensions of the equivalence are compatible: they can be 
combined. To list the extensions in the order given: the perdurantist can 
consider a {\em non-eternal} object, conceiving it in terms of {\em 
instantaneous stages} (so using functions $q_t$, with ${\rm dom}(q_t) = 
\{t\}$), locating it in {\em spacetime} $\cal M$ rather than space, and taking 
it as spatially {\em extended}: so that $q_t(t) \subset {\cal M}$. 

\subsection{The case of more than one object}\label{33:twoormore}
What about the case where two or more objects are in play? 
Again, the equivalence extends to this case; and it does so equally well for 
the various sub-cases---i.e. whether the objects are point-particles in a 
void, or spatial parts (including point-particles) of a common extended 
object, or distinct extended objects. From the endurantist perspective, the 
situation is straightforward. In all these sub-cases, the endurantist will 
have a collection of functions $q_i$, where $i$ now runs over an index set, 
which we can allow to be infinite and even uncountable. Just as before: each 
$q_i$ defines its restrictions $q_i \mid_{[a,b]}$, so that the endurantist can 
claim to recover the perdurantist's stages.

But from the perdurantist's perspective, the situation is more subtle. Yes, 
the equivalence extends to many objects, and does so equally well for the 
various sub-cases listed. But it is worth distinguishing two different ways 
the perdurantist might construct the endurantist's functions $q_i$. For the 
second brings out the limitations of {\em pointillisme}: that is, the second 
of these two perdurantist constructions {\em cannot} be adapted to using 
instantaneous stages. 

First, the perdurantist can work with a {\em doubly-indexed} collection of 
functions, $q_{i, [a,b]}$, which for each fixed $i$ meshes on overlapping 
values of the second index, in the sense of eq. \ref{eq;qsmesh}, so that for 
each fixed $i$ there is a unique function $q_i$ that yields the given $q_{i, 
[a,b]}$ by restriction---so that the perdurantist can claim to recover the 
endurantist's enduring object labelled $i$. In short: the equivalence of 
Sections \ref{31:fusingworldlines} and \ref{32:generalztns} carries over, the 
index $i$ just carrying along throughout (again: no matter how large the 
index-set). 

It is tempting to object to this construction that the perdurantist's  use of 
an index $i$ as  the first coordinate of a double-index amounts to 
presupposing the notion of persistence, in an illegitimate way. But I say 
`tempting' to indicate that the objection is not conclusive. For the 
dialectical situation here is not crystal-clear, since (as I admitted above) I 
have not tried to formulate endurantism and perdurantism precisely. In 
particular, I have not initially stated (for any of my equivalence's cases) 
exactly what resources each side is allowed to assume, nor what exactly is 
intended by ``recovering'' the perduring/enduring object of the other side. 
For example, a perdurantist  might reply to this objection with a {\em tu 
quoque}: saying that the endurantist's use, for the case of two or more 
objects, of the same index $i$ surely also presupposes persistence. I also 
agree that to this, some endurantist might reply that the endurantist's (but 
not the perdurantist's) presupposing persistence is legitimate, since the 
endurantist makes no claim to analyse persistence. And I agree that this 
rejoinder may be right, for some endurantists. And so it goes. As I said, the 
dialectical situation is unclear---and this paper will not aspire to getting 
it crystal-clear.  Here, it will be enough to discuss---in the following two 
Subsections---some issues about what resources each of the two sides might 
assume, and about what they can ``recover'' of the other side's notion of 
persistence.

\subsection{Avoiding double-indexing}\label{34:avoiddouble}
The first comment to make is that the perdurantist can make another 
construction, which ``recovers'' enduring objects, without assuming 
double-indexing from the beginning. More precisely, the perdurantist can do 
this provided that: (a) they do not work with instantaneous stages; and (b) 
the given enduring objects are mutually impenetrable, i.e. no two of them are 
located at the same place at the same time. (And if they {\em are} penetrable, 
the endurantists themselves arguably have work to do in regimenting or 
justifying their index $i$ ...) The idea of the construction is that 
impenetrability enables one to use an equation like eq. \ref{eq;qsmesh} to 
build up whole worldlines uniquely in the same way as before, i.e. by fusing 
the worldlines of stages---even though the stages are now {\em not} given as 
stages of one object rather than another (i.e. are not given using an index 
$i$).

     Let us take the case where the endurantist's enduring impenetrable 
objects are each of them eternal and are labelled by $i$, which runs over an 
index set $I$; and where the perdurantist is to ``recover'' them using stages 
corresponding (for each object) to all closed intervals $[a,b] \subset \mathR$ 
of time that are of finite length; i.e. $b$ is not equal to $a$ (no 
instantaneous stages). To do so, the perdurantist needs only to assert the 
existence of a singly-indexed family $\cal F$ of functions $q_{\al}$ ($\al$ 
running over {\em some} index-set) with the  five properties, (i)-(v) below. 
The first and second properties concern domains and codomains; the third 
expresses that there are enough perduring objects to recover all the 
endurantist's set of objects (a set whose size is the size of $I$); the fourth 
expresses the eternity of the objects; the fifth property  is the main one, 
expressing the impenetrability assumption.

The properties (i)-(v) are:---\\
\indent (i): for all $q_{\al}$: ${\rm dom}(q_{\al})$ is a closed interval 
$[a,b] \subset \mathR$, with $b$ not equal to $a$. \\
\indent (ii): for all $q_{\al}$: the values of $q_{\al}$ are points in space, 
say $\mathR^3$, or are points in spacetime $\cal M$, or are subsets of space 
or of   spacetime---according as the objects considered are located in space 
or spacetime, and are or are not spatially unextended. \\
\indent (iii): for each closed interval $[a,b] \subset \mathR$, the family 
$\cal F$ contains as many functions $q_{\al}$ with ${\rm dom}(q_{\al}) = 
[a,b]$ as there are elements in the (endurantist's) index set $I$.\\
\indent (iv): for any two overlapping closed intervals $[a,b], [c,d]$: for any 
function $q_{\al_1}$ with ${\rm dom}(q_{\al_1}) = [a,b]$, there is a function 
$q_{\al_2}$ with ${\rm dom}(q_{\al_2}) = [c,d]$ which agrees with $q_{\al_1}$ 
on the overlap of their domains, i.e. eq. \ref{eq;qsmesh} holds.\\
\indent (v): for any two overlapping closed intervals $[a,b], [c,d]$; for any 
$t \in [a,b] \cap [c,d]$; and for any two functions $q_{\al_1}, q_{\al_2}$ 
with ${\rm dom}(q_{\al_1}) = [a,b], \; {\rm dom}(q_{\al_2}) = [c,d]$:
\be
\mbox{If} \;\; q_{\al_1}(t) = q_{\al_2}(t) \;\; , \;\; \mbox{then} \;\;
\forall t' \in [a,b] \cap [c,d] \;\;\;\; q_{\al_1}(t') = q_{\al_2}(t') \;\; .
\label{eq;meshwithoutdoubleindex} 
\ee

To sum up:--- Property (v) says, in words: two functions representing stages, 
that have overlapping domains of definition, and agree on some argument $t$ in 
that overlap, must agree throughout that overlap, in the sense of eq. 
\ref{eq;qsmesh}. This agreement reflects the fact that the two functions 
represent overlapping stages of a single  persisting object. Taking (iv) and 
(v) together: (iv) states the existence of a continuation of any stage into 
the future (even the distant future, by letting $d >> b$); (v) makes that 
continuation unique.

So given such a family $\cal F$, the perdurantist can construct the worldlines 
of persisting objects in much the same way as they did for a single object. 
Starting with any function $q_{\al}$, with domain $[a,b]$ say, representing 
the $[a,b]$-stage of some object, condition (iv) implies the existence of a 
continuation upto any time $d$ in the future; (v) makes that continuation 
unique; and condition (iii) ensures that by considering all the functions 
$q_{\al}$ with domain $[a,b]$, we can recover all the endurantist's objects 
indexed by the index set $I$.

For the sake of completeness, I shall spell  out the argument of the last 
paragraph, showing how to introduce double-indexing for  
 $\cal F$, with the index $i$ given by the objects' non-overlapping locations 
on some fiducial time-slice. (But the details are not needed later and can be 
skipped.) There are three steps.\\
\indent (1): Pick any $[a,b]$ and any $t \in [a,b]$. Let any $q_{\al_1}$ with 
dom($q_{\al_1}$) = $[a,b]$ be given two indices:  i) $q_{\al_1}(t)$ and ii) 
its domain, $[a,b]$.\\
\indent (2): Now consider any other function $q_{\al_2}$ with dom($q_{\al_2}$) 
= $[c,d]$ say. There is a chain of intervals from $[a,b]$ to $[c,d]$, with 
adjacent intervals overlapping, i.e. $[a_1,b_1] := [a,b]$, $[a_2,b_2]$ with 
$[a_1,b_1] \cap [a_2,b_2] \neq \emptyset$ etc ... $[a_n,b_n] := [c,d]$. (In 
fact, by (iii) we can assume the chain has just three members: no matter how 
far apart the intervals $[a,b]$ and $[c,d]$ are, the interval between 
$\frac{a+b}{2}$ and $\frac{c+d}{2}$ overlaps both of them.) By (iii)-(v), this 
chain determines a unique function $q'$ with dom($q'$) = $[a,b]$, whose unique 
continuation, defined through the chain (in the sense of properties (iv) and 
(v)), on the interval $[c,d]$ is the {\em given} $q_{\al_2}$.\\
\indent (3): Now we label $q_{\al_2}$ by: i) the index $q'(t)$ given in step 
(1) above to $q'$ and ii) $q_{\al_2}$'s own domain $[c,d]$. Thus $\cal F$ 
becomes double-indexed in the desired way. 

Again, the construction can be varied and generalized in obvious ways: e.g. 
one could allow for non-eternal objects, and the perdurantist  could  work 
with open rather than closed intervals of times. But more important for us:  
the construction does not work if we use only instantaneous stages---which 
shows a limitation of {\em pointillisme} ...

\subsection{Trouble for {\em pointillisme}}\label{35:troublepism}
Recall from the end of Section \ref{sec;perseintrCush} that {\em pointillisme} 
vetoes temporally extrinsic properties as well as spatially extrinsic ones; so 
that a {\em pointilliste} perdurantist  will seek to analyse persistence as a 
matter of some suitable relations between instantaneous stages. So let us 
consider trying to revise the above five conditions so as to use only 
instantaneous stages. 

Agreed, conditions (i)-(iii) cause no problem. For these, the revision amounts 
to setting $b := a$ so as to make the closed interval $[a,b]$ degenerate into 
the singleton set $\{a\}$; and the conditions then express that at all times, 
there are as many instantaneous stages (spatially unextended or extended, as 
the case may be, according to (ii)) as there are objects in the endurantist's 
index-set $I$. But similar revisions of conditions (iv) and (v) make them 
vacuously true, since no two distinct singletons $\{a\}, \{c\}$ overlap; and 
being vacuously true, (iv) and (v) no longer state the existence and 
uniqueness of continuations of a given stage. 

Presumably, the perdurantist could impose some further assumptions so as to be 
able to recover from instantaneous stages the endurantist's many objects,  
without reverting to Section \ref{33:twoormore}'s  assumption of a 
doubly-indexed family. But since these assumptions are liable to be contested, 
I conclude that perdurantism is liable to face trouble if it is {\em 
pointilliste}, i.e. tries to work only with instantaneous stages. Certainly, 
{\em aficionados} of the rotating discs argument will  recognize the failure 
of Section \ref{34:avoiddouble}'s  construction when applied to instantaneous 
stages as a formal expression of the idea of that argument---that in a 
homogeneous disc, the timelike lines of qualitative similarity run ``every 
which way''.

 There is  one main exception to this failure: i.e. one salient further 
assumption that presumably enables the perdurantist  to recover from 
instantaneous stages the endurantist's objects. Namely, the assumption that 
the objects consist of point-particles separated from one another by empty 
space (or by a fluid made of some different kind of matter). In this special 
case, it seems the perdurantist can manage with just instantaneous stages. 
Namely, they can  recover persisting particles by tracking curves of 
qualitative similarity; or alternatively, curves of the occupation of 
spacetime  points by matter (or by matter of the particles', rather than the  
fluid's, kind). I will  examine this idea, and defend a precise form of it, 
from Section \ref{ssec;reconstructCush} onwards.

To sum up this Section, I have shown:\\
\indent (a): a formal equivalence of endurantism and perdurantism, based on 
the idea that a function fixes and is fixed by the set of its restrictions to 
subsets of its domain; and\\
\indent (b) how the equivalence  fails under certain {\em pointilliste} 
assumptions (viz. avoiding double-indexing, using instantaneous stages, and 
assuming continua not point-particles).

\section{Keeping Track of Particles}\label{ppsCush}
I turn to the topic of  criteria of identity over time; (also called 
`diachronic criteria of identity'). In Section \ref{needcoiCush}, I discuss 
this in purely philosophical terms, distinguishing various senses of 
`criterion of identity' and arguing that the endurantist  and perdurantist 
face similar questions about such criteria---to which they can give  similar 
answers. In Section \ref{ssec;reconstructCush}, I specialize to particles, and 
to what Section \ref{needcoiCush} calls their `epistemic criteria of 
identity': viz. our grounds or warrants for judgments that a given particle at 
one time is the same persisting particle (whether enduring or perduring) as a 
given particle at another time. The idea of such criteria prompts some 
comparatively precise questions about what bodies of information are 
sufficient to justify such a judgment. The next two Subsections 
(\ref{sssec;negresultCush} and \ref{sssec;posresultCush}) address some of 
these questions. The main result, using the Heine-Borel theorem to give a 
simple formal model of how such judgments can be justified, is in Section 
\ref{sssec;posresultCush}. Finally, I  discuss the prospects for other results 
(Section \ref{sssec;discussionCush}).

\subsection{Criteria of identity: variety and agreement}\label{needcoiCush}
In this Section, I make some general comments about: (1) the variety of 
notions that `criterion of identity' covers (Section \ref{sssec;varietyCush}); 
and (2) the properties invoked by criteria of identity (Section 
\ref{sssec;agreeCush}). These comments are intended to  report a consensus. I 
believe that they are largely independent of the endurantism-perdurantism 
debate; and in particular, that endurantism and perdurantism (and the mixed 
view mentioned in (ii) at the end of Section \ref{sec;perseintrCush}) face 
some common questions about criteria of identity, and can often give  the 
same, or similar, answers to them. (Later Subsections  will support this claim 
as regards the objects of classical mechanics.)

One general reason for this independence is worth stressing at the outset: 
namely, that it is a mistake to think that only the perdurantist owes an 
account of persistence. That is, there is a tendency to think that while the 
perdurantist certainly owes an account (in the jargon: a definition of the 
genidentity  relation among temporal parts), the endurantist does not, since 
for them ``persistence is just good old identity''. This is wrong, on two 
counts.\\
\indent First: for {\em both} sides of the debate, persistence involves 
identity. For perdurantists say that a rock, that has temporal parts at noon 
and 12.05, exists at noon and also at 12.05. It is identical with itself. So, 
also for the perdurantist,  something that exists at noon is identical with  
something that exists at 12.05; (this point is made by e.g. Sider (2001, pp. 
54-5)).\\
\indent  Second: all parties need to provide criteria of identity for objects, 
presumably invoking the usual notions of qualitative similarity and-or 
causation (cf.  Sections \ref{sssec;varietyCush} and \ref{sssec;agreeCush}). 
In particular, an endurantist who denied the need to do so would thereby allow 
that their ``good old identity'' could in principle  come  apart from any 
criterion invoking such notions, no matter how plausible, precise, suitably 
restricted etc. it was.  Which is an odd, perhaps even unintelligible, idea. 
It would  mean that for some persisting object $o$, (i) one's best (most 
plausible etc.) criterion ruled that $o$ at time $t$ is the same persisting 
object as $o'$ at $t'$, and yet (ii) this endurantist said that  in fact $o$ 
at $t'$ is not $o'$ at $t'$---it is somewhere else, or no longer exists.

\subsubsection{The variety of notions}\label{sssec;varietyCush}
`Criterion of identity' covers a variety of notions, which we can broadly 
distinguish in terms of two contrasts, which I will call: (i) ontic-epistemic, 
and (ii) conceptual-empirical. (As the generality of these labels suggest, 
these contrasts apply to many other topics in philosophy besides persistence 
and criteria of identity.) I think that by and large, philosophers' usage 
favours the first of each pair; i.e. `criterion of identity' tends to be taken 
as ontic and conceptual. But  my discussion in later Subsections will 
concentrate on epistemic and empirical criteria, for particles in classical 
mechanics.  

\indent (i): {\em The ontic-epistemic contrast}.  All parties agree that there 
is a contrast between persistence, and the grounds or warrants we typically 
use to make judgements of persistence. The obvious everyday example is people. 
We recognize them by their faces. But we all admit that this is a short-cut: 
indeed, in two senses. Not only is it fallible in the everyday sense: `Hello! 
... Oh, I'm sorry: in the poor light, I thought you were someone I knew'. It 
is also fallible even when one gets the {\em face} right. In a court-room 
drama, guilty A has had plastic surgery so as to look just like innocent B 
used to look, while B's face has changed radically since witness C knew them; 
so that C, who thought they saw B at the scene of the crime, got the face {\em 
right}---but got the person wrong. Nor is it just face, and similarly 
ubiquitous and common-sense grounds for persistence judgments, that are 
fallible in this second sense. So are special, technical grounds. In a 
science-fiction court-room drama, guilty A is a clone of innocent B, so that 
both the witness using face and the detective using DNA-tests think that B was 
at the scene of the crime. Just as for people, so also for other objects, 
natural and artificial, such as rocks and chairs: we naturally distinguish 
between the fact of persistence, and our grounds---everyday and technical, 
occasional or systematic---for judgments of persistence.\\
\indent By and large, `criterion of identity' tends to be used for the former, 
ontic, notion: in philosophers' jargon, the `constitutive facts' of 
persistence---and perhaps for some `canonical' or `analytically correct' 
grounds for judgment of persistence. So, to take a simple example: if the 
criterion of identity of some solid object is given by sameness of constituent 
matter, the canonical grounds for judgment might be that one has tracked all 
that matter continuously in space and time.\footnote{Agreed, this example is 
controversial, not least because philosophers recognize that the idea of 
tracking matter is problematic (e.g. Robinson 1982): a topic to which I return 
in Section \ref{ssec;reconstructCush}. But  the controversy is no bar to the 
example's present role.} But my later discussion (Section \ref{sssec;making 
precise} onwards) will concentrate on the epistemic notion: in this example, 
the notion of {\em how}  one could  track matter. 

(ii): {\em The conceptual-empirical contrast}. All parties also agree that the 
topic of criteria of identity---like, I daresay, almost any topic in analytic 
philosophy---can  be approached in either of two ways:---\\
\indent  (a): As a field for conceptual analysis: one analyses non-technical 
concepts, and relates them to each other, aiming to describe but not to revise 
those concepts.\\
\indent (b): As a field for constructing the best theory: where one can appeal 
to scientific  technicalities, and `best' can include requiring empirical 
adequacy; an enterprise that could be revisionary, rather than descriptive, of 
the original concepts.\\ 
\indent Of course, each of (a) and (b) is a broad church. For example, some 
practitioners of (a) abandon the traditional requirement that analyses---i.e. 
in this discussion, criteria of identity---be finitely stated, and aim only to 
provide a supervenience basis, so that ``infinitely long analyses'' are 
allowed.  Besides, there is obviously a spectrum from (a), through appeals to 
common sense knowledge (``folk science''), to (b)'s appeals to technical 
science.  `Criterion of identity' tends to be used for the conceptual analysis 
end of this spectrum. But my later discussion (Section \ref{sssec;making 
precise} onwards) will concentrate on the empirical end. 

\subsubsection{Agreement on the properties invoked}\label{sssec;agreeCush}
Many, I daresay most, philosophers agree that for most objects, their 
criterion of identity will invoke one or both of the following two factors 
(which also might well overlap): qualitative similarity, and causal 
relatedness.
  
\indent Qualitative similarity concerns whether the object at the two times 
(or in perdurantist terms: the two stages) has suitably similar qualitative 
properties. Here, `suitably similar' is to be read flexibly. It is to allow 
for:\\
\indent (i) only a tiny minority of properties counting in the comparison;\\
\indent (ii) considerable change in the object's properties, provided the 
change is ``suitably continuous''; i.e. provided the object goes through some 
kind of chain of small changes.

\indent Causal relatedness concerns whether the state of the object at the 
later time (or the later stage) is suitably causally related by the earlier 
state or stage. Here again `suitably causally related' is to be read flexibly. 
It is to allow for:\\
\indent (i) various rival doctrines about causation---including a special 
variety of causation, called `immanent causation', that some philosophers 
believe relates an object at two times (or in perdurantist terms: relates its 
stages); (Zimmermann 1997);\\
\indent (ii) a suitable chain of states or stages linked by causation.

 However, most philosophers will also agree that  it is very difficult to go 
beyond this vague consensus to give precise  criteria of identity, i.e. 
precise necessary and-or sufficient conditions for persistence. And this is 
so, whether we take the criteria we seek to be ontic or epistemic: and whether 
we take them to be empirical or conceptual; and whether we take them to be for 
all objects in a very wide class, or for all objects in a narrow class, e.g. a 
given species of animal (or even narrower). Even if  we take what we believe 
to be the easier  option in these three regards (which is I suppose the 
second, in each regard), there are intractable issues.

The most obvious issue is about weighing competing factors. This occurs even 
if we consider only qualitative similarity, i.e. if we eschew causation. For 
example: even for our judgments about the persistence of  chairs, it is very 
hard to suitably weigh the various possible changes of properties, including 
functional properties, and changes of constituent matter. The situation is 
similar, and no doubt worse, for persons: here we need to suitably weigh 
psychological vs. bodily considerations. These examples raise the topic of 
vagueness. We can no doubt all agree that which (if any) of tomorrow's objects 
counts as today's chair or person can ultimately be a vague matter---but how 
exactly should we think of that? The philosophical literature addresses this 
issue in detail. In fact, I believe the balance of evidence favours 
perdurantism, but I will not argue this here.

Also, many philosophers (I amongst them) will be sceptical about  appealing to 
causation---the general notion, not just the idea of a special variety, 
immanent causation. It is not just that causation seems too controversial and 
ill-understood to be a central notion in  criteria of identity. Also, a good 
case can be made that there is no single causal relation, so that a 
philosopher who appeals to causation for criteria of identity needs to choose 
one of the range. (Hitchcock makes such a case; more specifically, he 
advocates a pluralism about causation along two dimensions (2003, especially 
pp. 5-9).) Another threat is that some accounts of causation (e.g. Dowe 2000) 
deliberately presuppose the notion of persistence---threatening a logical 
circle.\footnote{A brief look at the literature in analytic metaphysics 
reveals a wider phenomenon of intellectual compartmentalization, which is 
worrying though of course  undertandable: viz. metaphysical accounts of 
various topics invoke causal notions but typically do not reflect the agonies 
which philosophers of science have gone through to give accounts of 
causation.}

But of course, this is  not the place to explore these intractable issues. 
Fortunately, they are in any case  independent to a large extent of the 
endurantism-perdurantism debate: as witness the fact that the above consensus 
about which properties to invoke is common to endurantists, perdurantists (and 
advocates of the mixed view)---my formulations of the two factors, qualitative 
similarity and causal relatedness, does not favour endurantism or 
perdurantism.\footnote{Cf. the general reason for this independence at the 
start of this Subsection: the fact that for the endurantist, persistence is 
``good old identity''  does not mean they need not address, along with other 
philosophers, the usual puzzle cases about persistence, e.g. about personal 
identity.}

\subsection{Particles}\label{ssec;reconstructCush}
From now on I specialize to particles, as understood in classical mechanics: 
i.e. either point-particles, or bodies small and rigid enough to be modelled 
as point-particles (in which case, the discussion is silent about criteria of 
identity for the bodies' spatial parts). I will assess the widespread idea 
that particles'  criterion of identity is given  by what Section 
\ref{IntrCush} called `{\em Follow}': i.e. by following the timelike curves of 
maximum qualitative similarity. I will first briefly defend this as an ontic 
criterion (Section \ref{sssec;doubtscommon}); then I will develop it in much 
more detail as an epistemic criterion.

\subsubsection{The ontic criterion}\label{sssec;doubtscommon}
For particles that move in a void (or in a continuous  fluid made of a 
different kind of matter), and are each assumed to have a continuous 
worldline, the idea of {\em Follow} seems plausible as an ontic criterion of 
identity. The precise proposal will be along these lines: a particle $o$ at 
time $t$ is the same particle as $o'$ at time $t'$ iff the unique continuous  
timelike curve of maximum qualitative similarity through the spatial location 
of $o$ at $t$ passes through the location of  $o'$ at time $t'$. But we need 
to define `curve of maximum qualitative similarity', and maybe make ancillary 
assumptions, in such a way as to secure a unique such curve. 

Let us first assume that the particles do not collide. Agreed, that is a big 
assumption. But it is   an endemic one in mechanics, since collisions of 
point-particles are both kinematically and dynamically intractable; e.g. under 
Newtonian gravity, two colliding point-particles each have infinite kinetic 
energy at the instant of collision.  Let us also allow particles to change 
their qualitative properties: but only continuously as a function of time; and 
never in such a way as to be indistinguishable from any surrounding 
fluid---some ``charge'' or ``colour'' must remain different from the fluid's. 
These assumptions mean that, without worrying about how exactly to define 
`qualitative properties' (which is philosophically very problematic), we can 
be confident that any such definition will enable us:\\
\indent (i) to define through the spatial location of each particle  at each 
time a unique continuous  timelike curve of `maximum qualitative similarity' 
(where this phrase will be spelt out in terms of comparisons of properties at 
arbitrarily close times);\\
\indent (ii) and so to contend that this curve is the worldline (initially 
assumed continuous) of the given particle.

Unsurprisingly, one can vary these assumptions somewhat.\\
\indent (a): Subject to a {\em proviso}, one can allow for discontinuous 
changes of properties, and still define the  continuous curves of qualitative 
similarity as intended (i.e. still recover the worldlines of particles). 
Roughly speaking, the {\em proviso} should  say that any discontinuous changes 
at a time $t$, as a result of which particles $o, o'$ become more like one 
another than like their previous selves,  are outweighed in the assessment of 
similarity by other properties that $o$ and $o'$ are not ``exchanging'' at 
$t$. (Nevermind the details, which would require assumptions about such 
intractable issues as weighing competing properties; cf. the end of Section 
\ref{sssec;agreeCush}).\\
\indent (b): More important for us: for particles in a void, it seems that 
instead of considering all their qualitative properties, we can just consider 
one quality of spacetime points: viz. being occupied by matter. For with $n$ 
point-particles in a void that are assumed not to collide, there is through 
each spacetime point occupied by matter, just one continuous  timelike curve 
along which  the quality of being occupied by matter is maintained: viz. the 
particle's worldline.

I said that this ontic criterion was plausible. Before considering analogous 
epistemic criteria, I should address two criticisms of it. The first is in 
effect that the criterion is too weak, and the second that it is too strong. I 
shall have more sympathy for the second criticism; but will still defend the 
criterion, for particles assumed to have continuous worldlines.

(1) You might object that the criterion is too weak. That is, in terms of 
defining genidentity in terms of qualitative similarity: you might object that  
the {\em definiens} is not sufficient for  genidentity. Thus some philosophers 
fantasize that a god could instantaneously destroy an object and replace it 
immediately with a qualitative replica: suggesting that a continuous timelike 
curve  of qualitative similarity  is not sufficient for persistence.\\
\indent To reply to this objection, it would not help to have the criterion 
require qualitative matching, instead of just similarity: for the qualitative 
properties of the destroyed object and its replica need never change. Nor 
would it help to endorse (b) above, and have the criterion  follow occupation 
by matter rather than other qualitative properties. For the criterion cannot 
follow occupation by the {\em same} matter, on pain of presupposing 
persistence: and following occupation by  {\em some} matter faces the original 
objection just as much as following qualitative similarity.\\
\indent Rather, what seems to be missing in such a case is an appropriate 
causal relation between the destroyed object's state just before destruction, 
and the replica's state just afterwards. Hence these philosophers conclude 
that criteria of identity should invoke causation: and perhaps the special 
variety, immanent causation, that is meant to relate an object at two times 
(or in perdurantist terms: its stages). (For references to these replacement 
fantasies, cf. e.g. Zimmermann 1997 p. 435-437).

\indent (2): On the other hand, you might object that the criterion is too 
strong, i.e. the {\em definiens} is not necessary to genidentity. Here the 
idea is to question the criterion's initial assumption that each particle has 
a continuous worldline. Surely a persisting  particle could jump about 
discontinuously? Though this is forbidden by classical 
dynamics,\footnote{Besides, it is notoriously too simplistic a way to think 
about quantum dynamics---even apart from Cushing's favoured pilot-wave 
theory!} it seems logically possible. Indeed,  we could gather evidence for 
it, by finding appropriate  patterns in the changing properties of the 
particle. To take a simple example: suppose we find that the particle that 
seems to sometimes jump discontinuously in space (say a metre every second) 
cools down, like ordinary objects do---and has equal temperatures just before 
and just after a jump. That would confirm the idea that the particle persists 
across jumps.

Both these objections, (1) and (2), will strike most physicists as a mere  
parlour-game, since both give no physical  details (nor other details) about 
their main idea, destruction and jumping respectively. Fair comment, say I. 
But note that, at least on the conceptual analysis approach to criteria of 
identity (Section \ref{sssec;varietyCush}),  this kind of objection, that 
describes without empirical detail a ``bare'' logical possibility, can be 
enough to refute a proposed analysis.\footnote{Besides, it seems only fair to 
the philosophers to point out that so great a physicist as John Bell regarded  
an idea even odder than (2)'s jumping about as the best construal of the 
Everett interpretation of quantum theory! The idea is that systems should jump 
discontinuously between states, which include  records of the past, in such a 
way that the jumping is undetectable. Cf. Bell (1976, p. 95; 1981, p. 136); 
Butterfield (2002, p. 312-316) is a discussion.}

 So, ``playing the game'' of conceptual analysis, for what it is worth: as to 
(1), I have already said that I am suspicious of appeals to causation (Section 
\ref{sssec;agreeCush}). Indeed I think that one can bite the bullet about this 
objection; or at least, a perdurantist can. That is: one can maintain that the 
{\em definiens is} sufficient for genidentity. So in a world of the sort 
described, the  object instantaneously destroyed and the immediately 
succeeding replica would be the same perduring object.\footnote{I agree that 
perhaps  the endurantist will say that the very word `destruction' forbids the 
replica being the same persisting object.} Here, I should sugar the pill of 
this counterintuitive verdict, by making a  point which also applies in many 
similar cases of conceptual analysis. Namely: this is put forward as the 
verdict given by {\em our} concept of persistence when used to describe {\em 
that} world; we can admit that inhabitants of that world may well have good 
reason to use another concept that denies persistence in such a case, say 
because a systematic pattern in the god's destructive acts makes it important 
to sharply distinguish an object before and after the act.

As to (2), I am more sympathetic. Such jumping about is surely logically 
possible, and appropriate  patterns in particles' properties  could give us 
evidence of it. But to write down a {\em definiens} for genidentity, i.e. an 
analysis of persistence, that accommodated the possibility would be a tall 
order: we would again have to address such intractable issues as weighing 
competing properties (again cf. the end of Section \ref{sssec;agreeCush}). I 
shall duck out of this, and simply take the ontic criterion as intended for 
particles with  continuous worldlines. That is fair enough, in so far as the 
criterion assumed such worldlines at the outset: so the idea of jumping about 
represents, not so much an objection to the criterion, as a limitation of it. 
In short, I think we should ``divide and rule'': if we agree that jumping 
about is possible, then jumping particles fall outside the scope of the ontic 
criterion.

\subsubsection{Epistemic criteria}\label{sssec;making precise}
So much by way of defending 
the idea of following timelike curves of qualitative similarity (Section 
\ref{IntrCush}'s `{\em Follow}') as an ontic criterion, at least for particles 
assumed to have continuous worldlines. From now on, I will discuss this idea 
as providing  epistemic criteria for such particles. So I ask: what bodies of 
information about ``tracking trajectories'' are our grounds  for judgments 
that a given particle at one time is the same persisting particle (whether 
enduring or perduring) as a given particle at another time? In this Subsection 
I formulate  some more precise versions of this question. Then subsequent 
Subsections will answer some of these versions.

To make vivid at the outset what I intend by an epistemic criterion (ground 
for judgment of persistence), it helps to see how the ontic criterion would be 
of no use to an engineer trying to design a robotic vision system which tracks 
$n$ particles (treated as point-particles) moving continuously through  space.  
Imagine that the particles are indistinguishable (i.e. not distinguished by 
colour or any other property to which the robot's eye is sensitive); and that 
the robot's eye delivers to the central processor a discrete-time sequence of 
instantaneous configurations of the $n$ particles, the configurations being 
reported in terms of particles' coordinates in a chosen cartesian coordinate 
system.  Since the particles are indistinguishable,  each instantaneous  
configuration is an {\em unordered} set of $n$ 3-tuples (ordered triples) of 
real numbers, each 3-tuple giving the coordinates of one of the particles. So 
each configuration is a set $\{q_1,\dots,q_n\}, \; q_i \in \mathR^3$, where 
the label $i$ has {\em no}  significance in common among different sets. (Here 
and from now on, $q_i$ etc. will represent a spatial location, not a 
spatiotemporal one: as they often did in Section \ref{intertransCush}.)\\
\indent The task of the central processor is to determine for any choice of 
particle in each of two configurations, whether the choice is of the same 
particle. That is: is the particle with coordinates $q_i$ in a configuration   
$\{q_1,\dots,q_n\}$ the same as (genidentical with)  the particle with 
coordinates $q'_j$ in some other configuration   $\{q'_1,\dots,q'_n\}$?\\
\indent The engineer designing such a processor will not thank you for telling 
her that the answer is Yes iff there is a continuous timelike curve of 
qualitative similarity (or matter-occupation) from $q_i$ at the first 
configuration's time to $q'_j$ at the second's. She knew {\em that} already!

At this point, physicists will suggest that the general topic here is their 
traditional prime concern: solving the equations of motion of some given 
dynamics. This prompts three comments, in descending order of importance:\\
\indent (1): First, I reply: fair comment. But I want to pose the search for 
epistemic criteria at a different level than this suggestion: a level that is 
in some  ways more general, e.g. in that no laws of dynamics are assumed, but 
in other ways more specific, e.g. in focussing only on judgments of 
persistence. (My level will be akin to the engineers' level, in that for them 
the system's dynamics is often {\em not} ``given''---it is unknown or 
intractable---so that to solve their problem, they have to exploit 
system-specfic details.)\\
\indent (2): Beware of an ambiguity: metaphysicians and physicists tend to use 
the word `determine', especially in  phrases like `determine the worldline' 
(of a given object) in different senses. Physicists naturally take this to 
mean solving the equations of motion so as to calculate the object's future 
position, given its initial position and velocity, and the forces acting on 
it. For metaphysicians, on the other hand, `determine' usually means 
`subvene'; so that `determining the worldline' is a matter of stating those 
facts on which the persistence of the object supervenes. Besides, this is 
often interpreted in terms of a finitely stateable criterion of identity at 
the ontic and conceptual (not epistemic and empirical) ends of Section 
\ref{sssec;varietyCush}'s two contrasts.\\
\indent (3): The ambiguity discussed in (2) is well illustrated by a point 
about velocity: that the notion of velocity presupposes the persistence of the 
object concerned. For average velocity is a quotient, whose numerator must be 
the distance traversed by the given persisting object: otherwise you could 
give me a superluminal velocity by dividing the distance between me and the 
Sun by a time less than eight minutes. So presumably, average velocity's 
limit, instantaneous velocity, also presupposes  persistence. Accordingly, 
metaphysicians often say that instantaneous velocity cannot, on pain of 
circularity, be in the supervenience basis for facts about 
persistence.\footnote{However, I argue in (2004, 2004a) that once we reject 
{\em pointillisme}, we can in a sense admit velocity in to the supervenience 
basis.} But on the other hand, in a classical (Newtonian not Aristotelian!) 
dynamics, initial velocity (or momentum) is an essential part of the initial 
conditions that `determine the worldline'.

There are of course  various ways to make precise the idea of an epistemic 
criterion that tracks the worldlines of particles: as I put it above, various 
ways to choose a level. I shall adopt a simple and abstract level, ducking out 
of making contact with the details of a real engineering problem! To be 
specific, I begin by making the following assumptions. But it will be obvious 
from the results in subsequent Subsections how several of these, e.g. the 
assumption of Euclidean geometry, can be weakened.

I will  assume a (relativistic or non-relativistic) spacetime manifold $\cal 
M$ which---at least in the spacetime region with which we are concerned---can 
be foliated into instantaneous i.e. spacelike slices, and covered by a 
timelike congruence of curves that we think of as persisting spatial points. 
Furthermore, I assume that the slices can be labelled by numbers in the real 
interval $[a,b] \subset \mathR$, and that the points have a Euclidean 
geometry, so that I can represent the distance between them in cartesian 
coordinates in the usual way. More specifically, I shall consider a closed  
temporal interval (slab of spacetime) $T$ coordinatized as $T = [a,b] \times 
\mathR^3$.

I represent the $n$ particles as point-particles with continuous worldlines. 
So I assume there is a  set of $n$ continuous  timelike curves $\gamma_i: 
[0,1] \subset \mathR \rightarrow {\cal M}, i = 1,\dots,n$, each of which 
registers throughout $T$  in the obvious sense that for all $i$, the 
worldline, i.e. the image (range) of $\gamma_i$, intersects (just once) each 
hypersurface of $T$, $\{t\} \times \mathR^3, \; t \in [a,b]$. I assume also 
that the particles never collide in $T$, i.e. $T \cap \left(\cap_i {\rm 
Ran}(\gamma_i)\right) = \emptyset$. So, turning to the spatial location, 
rather than spatiotemporal location, of particle $i$, I write (the image of) 
$\gamma_i$ on $T$ as $\gamma_i(t) = (t, q_i(t)) \in T$ with $q_i(t) \in 
\mathR^3$. So  an instantaneous configuration of all $n$ particles is given by  
a set of distinct points in $\mathR^3$:  $\{q_1,\dots,q_n\}, \; q_i \in 
\mathR^3$. 

With these assumptions in place, I now ask what bodies of information are 
sufficient to answer the question above, that the robot's central processor is 
to address. That is, now taking the label  $i$ to have {\em no}  significance 
in common among different configurations, the question: is the particle with 
coordinates $q_i$ in a configuration   $\{q_1,\dots,q_n\}$ the same as 
(genidentical with) the particle with coordinates $q'_j$ in another 
configuration   $\{q'_1,\dots,q'_n\}$?\footnote{Incidentally: these 
assumptions bring out that my project is a qualitative ``opposite'' to the 
Machian dynamical theories  of Barbour et al. (cf. Butterfield 2002 for 
references). In short: these theories  assume the notion of persistence (and 
in their present form, some spatial structure such as a Euclidean geometry), 
and apply Machian principles to define further structure, especially a 
temporal metric. But my project assumes {\em a priori} some simple spatial and 
temporal structure, and asks if we can then  define persistence.}

Broadly speaking, my answer will be:\\
\indent  (i): Examples show that if the bodies of information are limited in 
certain ways, then they are not sufficient---the robot's central processor 
cannot solve the problem. (The case of one spatial dimension will be an 
exception.) But\\
\indent (ii): Under some general conditions, there are sufficient bodies of 
information: the central processor can solve the problem. The idea of these 
conditions is that particles  move ``slowly enough'', so that two particles  
could not, between the times of the two configurations, swap which spatial 
neighbourhoods they are in. I shall develop this idea  formally, using some 
elementary real analysis (the Heine-Borel theorem).\\
\indent Details are in Sections \ref{sssec;negresultCush} and  
\ref{sssec;posresultCush} respectively.

\subsection{Examples}\label{sssec;negresultCush}
Suppose we are given a choice of particle in each of two configurations, say 
$q_i$ in a configuration  $\{q_1,\dots,q_n\}$ and $q'_j$ in configuration   
$\{q'_1,\dots,q'_n\}$, with $i$ and $j$ having no  significance in common 
among different configurations. We are asked to 
determine  whether the choices are of the same particle.  Under what 
conditions can we do so?

Suppose first that space is one-dimensional, and that particles are 
impenetrable. Then we can immediately determine whether the same particle was 
chosen. We just need to order each configuration by spatial position along the 
real line, so that with $\sigma$ and $\pi$ the permutations that yield such an  
order from our given arbitrary labellings, we have:
\be
q_{\sigma(1)} < q_{\sigma(2)} < \dots < q_{\sigma(n)} \;\; \mbox{and} \;\; 
q'_{\pi(1)} < q'_{\pi(2)} < \dots <  q'_{\pi(n)}.
\ee 
Then, thanks to particles' impenetrability, the particle $\sigma(i)$ is the 
same particle as $\pi(i)$, for all $i= 1,\dots, n$.

But if space has dimension at least two, this strategy fails. Indeed, it is 
easy to give examples showing that  the problem is insoluble (in various 
senses, corresponding to various assumptions about what the body of 
information can contain). That is: no body of information (subject to the 
assumptions) can answer the question, `was the same particle chosen?'. As one 
would expect, these examples have a certain symmetry that makes the Yes and No 
answers to this question equally well supported. I will give two such 
examples.

(1): Suppose that the body of information is to be formulated wholly in terms 
of the two configurations, $\{q_1,\dots,q_n\}$ and  $\{q'_1,\dots,q'_n\}$. (We 
will not need to be more precise than this.) Now suppose we are given two 
configurations $C_1, C_2$ of two particles  moving in two-dimensional space, 
in terms of cartesian coordinates $\langle x,y \rangle$ on $\mathR^2$: 
\be
C_1 := \{ \langle 1,0 \rangle, \langle -1,0 \rangle \} \;\; \mbox{and} \;\;
C_2 := \{ \langle 0,1 \rangle, \langle 0,-1 \rangle \}.
\ee
We can also imagine that we are told ``for free'' that: (i) both particles 
orbit the origin at constant radius 1 unit, with a common sense of rotation, 
and a common constant speed, so that they are always opposite each other; and 
(ii) $C_2$ is later than $C_1$ by a quarter-period. Still, we cannot tell 
whether\\
\indent (a): this is anti-clockwise rotation, so that (using the ``free'' 
extra information) the particle at $\langle 1,0 \rangle$ in $C_1$ is the same 
as the particle at $\langle 0,1 \rangle$ in $C_2$ (and the particle at 
$\langle -1,0 \rangle$ in $C_1$ is the same as the particle at $\langle 0,-1 
\rangle$ in $C_2$); or:\\
\indent (b) {\em vice versa}: i.e. this is clockwise rotation, so that  the 
particle at $\langle 1,0 \rangle$ in $C_1$ is the same as the particle at 
$\langle 0,-1 \rangle$ in $C_2$ (and the one at $\langle -1,0 \rangle$ in 
$C_1$ is the same as the one at $\langle 0,1 \rangle$ in $C_2$).

(2): The second example is very similar. Suppose again that we are told two 
particles orbit the origin at constant radius 1 unit, with a common sense of 
rotation, and a common constant speed, so that they are always opposite each 
other; and that the angular speed is $2\pi$ radians per second. The two 
configurations  we are given are two identical copies of $C_1 := \{ \langle 
1,0 \rangle, \langle -1,0 \rangle \}$. We are also told that the time interval 
between the configurations is greater than 0.25 seconds, but less than 1.25 
seconds. Then we cannot tell whether  the time-interval is 0.5 seconds, so 
that the two particles have exchanged positions between the two 
configurations, or is 1.0 second, so that the particles have not exchanged 
positions.

Examples like these prompt the idea that the obstacle to answering the 
question `was the same particle chosen?' is our lack of suitable information 
about: (i) other configurations, especially at intervening times (example 
(1)); and-or (ii) the times at which the given configurations occur (example 
(2)). Agreed, we cannot in general expect to reconstruct the temporal order of 
a set of configurations.\footnote{Just imagine a single particle at rest in 
space, at say $(x,y,z) \in \mathR^3$; so that all configurations, without time 
labels, are the same: $\{(x,y,z)\}$. From this continuously large set of 
identical configurations, ``not even God'' could reconstruct their correct 
temporal order. Indeed, there are $\aleph_1^{\aleph_1}$  bijections from the 
point-set $[a,b] \subset \mathR$ to itself. So there are that many ways to 
linearly order this set of configurations. Agreed: in less ``symmetric'' 
cases, the assumption that each particle  has a continuous worldline can help 
us temporally order a set of configurations.}  But there is no strong reason 
to require that we pose the question `was the same particle chosen?' without 
information about the times of configurations. Nor is there strong reason to 
veto information about other configurations. After all, recall the robot's 
central processor. We assumed that it received from the robot's eye a 
discrete-time sequence of configurations, i.e. a whole set of configurations, 
not just two. And the eye could be equipped with a  clock that labels each 
configuration with its time, before it is passed to the central processor.

So let us consider  the question `was the same particle chosen?', now thinking 
of ourselves as being given a discrete-time sequence of time-labelled 
configurations. Now, the assumption that each particle  has a continuous 
worldline apparently makes our problem soluble. That is: it seems that if the 
time-step in the discrete-time sequence of configurations is small enough, our 
question `was the same particle chosen?' {\em can} be answered completely 
reliably. For with a small enough time-step, no particle can have traversed so 
great a distance as to have swapped places with another. The next Subsection 
will take up this line of thought.

\subsection{Bounding distances and speeds}\label{sssec;posresultCush}
I will argue that the intuition at the end of Section 
\ref{sssec;negresultCush} can be vindicated.  But as I admitted in Section 
\ref{sssec;making precise}, I will adopt an abstract level, ducking out of the 
details of real engineering problems. In particular, I will not assume we are 
given a discrete-time sequence. Rather,  I will think of ourselves as being 
given the whole set of time-labelled configurations, and then ask how the 
whole set determines which particle is which,  across different 
configurations.

 Even when the problem is reformulated in this way, the intuition above holds 
good. That is: given a spatial point  occupied by a particle in one 
configuration, we can reconstruct the worldline through the point, by using 
the fact that, thanks to the finite speed of all particles (and the 
no-collision assumption), no {\em other} particle could be  very close to the  
point at times very close to the given time. The main reason we can do this, 
even without assuming {\em ab initio} that there is a discrete time-step that 
is ``small enough'', is that some facts of elementary real analysis in effect 
guarantee to us that there is such a time-step. Specifically, I shall use the  
facts that a continuous real function on a closed bounded real interval is 
bounded, attains its bounds, and is uniformly continuous. I shall apply these 
facts---which are all corollaries of the Heine-Borel theorem---to functions 
representing the spatial locations of, and distances between, particles. 

It will be clearest to proceed in three stages. (1): I will state and comment 
on the assumptions I make. (2): I will state the notion of time-labelled 
configuration I use. (3): I will show how these configurations determine 
worldlines.

\paragraph{(1) Assumptions} As at the end of Section \ref{sssec;making 
precise}, I assume that $n$ point-particles, labelled $1,\dots,i,\dots,n$:\\
\indent a): each register on every spacelike slice $\{t\} \times \mathR^3$ of 
some closed temporal interval (slab of spacetime) $T := [a,b] \times \mathR^3$ 
($[a,b] \subset \mathR$); and\\
\indent b): each have,  during $[a,b]$, a continuous worldline, the (image of) 
a continuous timelike curve  $\gamma_i: [a,b] \rightarrow T, i = 1,\dots,n$; 
with spatial location represented by writing $\gamma_i(t) = (t, q_i(t)) \in T$ 
with $q_i(t) \in \mathR^3$ and \\
\indent c): never collide in $T$, i.e. $T \cap \left(\cap_i {\rm 
Ran}(\gamma_i)\right) = \emptyset$.

Three comments on these assumptions, in ascending order of importance:---\\
\indent [i]: In (3) below, I shall discuss (but not rely on) strengthenings of 
b) according to which each function $q_i: t \mapsto q_i(t) \in \mathR^3$ 
giving spatial location is (i)  differentiable at all $t$ so that each 
particle has a velocity, or even (ii) continuously differentiable, so that 
each particle's velocity is continuous.\\  
\indent [ii]: For clarity, I have stated these assumptions in a stronger form 
than needed. It will be obvious that  the result in (3) below is independent 
of the following: $T$'s spacelike slices being isomorphic to $\mathR^3$, or to 
each other; the number of spatial dimensions; whether there is an absolute 
rest (i.e. non-dynamical timelike vector field fixing the spacetime's 
connection); whether simultaneity is Newtonian or relativistic. Besides, the 
result can be adapted to extended objects provided they are {\em treated as 
wholes}.\\
\indent [iii]: On the other hand, I {\em do} need assumptions that will give 
space and time enough structure so that I can apply the corollaries of the 
Heine-Borel theorem. But I will not explore what the weakest such structures 
might be: that would amount to a project in advanced analysis. Suffice it to 
say here that: I need to consider a closed interval of time $[a,b]$, not an 
open one; and I need to assume that space (i.e. the set of persisting spatial 
locations, but not necessarily $\mathR^3$) is a metric space.

\paragraph{(2) Configurations}  Instantaneous configurations are to be taken 
as time-labelled; as not presupposing any facts of persistence; and for 
simplicity, as presupposing absolute space. Accordingly, I define a 
configuration as an {\em unordered} set of $n$ (absolute spatial) locations, 
taken as say triples of real numbers, that are occupied at a given time $t \in 
[a,b]$; together with the  time-label $t$. (The  time-label may as well be 
taken as an element of the set along with the rest, as given the set we can 
immediately identify it, viz. as the only real-number member.) So the 
configuration at time $t$ is the unordered set:
\be
\{q_1(t), \dots, q_n(t), t\} = \{<x_1(t),y_1(t),z_1(t)>, \dots, 
<x_n(t),y_n(t),z_n(t)>, t\}.
\label{eq;defconfig}
\ee    
A history $H$ of the system of particles  during the time-interval $[a,b]$ is 
represented by $n$  functions $q_i$. These define a set ${\rm Config}(H)$ of 
instantaneous configurations of the form eq. \ref{eq;defconfig}.

\paragraph{(3) The result} I will now show how, given that a spatial location 
$<x_0,y_0,z_0>$ is occupied at $t_0 \in [a,b]$, ${\rm Config}(H)$ determines 
the worldline, the $q$-curve, through $<t_0,x_0,y_0,z_0> \; \in \; T$.

Note first that ${\rm Config}(H)$ fixes all the inter-particle distances as a 
function of time. Of course,  in order not to presuppose facts about 
persistence, we  must not think of these distances as encoded in functions for 
each pair $\{i,j\}, \;\; i,j = 1,\dots,n, \;\; i \neq j$, ${\rm dist}_{ij}: t 
\in \mathR \mapsto {\rm dist}_{ij}(t)$ := the distance at time $t$ between 
particles $i$ and $j$. Rather, at each time $t$, we can only label from 1 to 
$n$ the $n$ spatial points that are then occupied, {\em without} regard to 
particles' persistence. So  the particle labelling is arbitrary and 
$t$-dependent, $i(t), j(t)$ etc. Then we can define ${\rm dist}_{ij}(t) := 
{\rm dist}_{i(t),j(t)}(t)$ := the distance at time $t$ between particles 
labelled at $t$ as $i$ and $j$. Since the labels can ``jump about'' 
arbitrarily, ${\rm dist}_{ij}(t)$ is {\em not} continuous as a function from 
$[a,b]$ to $\mathR$.

 But we can obtain a continuous function by taking the minimum over all pairs. 
That is, we define
\be
d(t) \;\; := \;\; {\rm min}_{{\rm pairs}\; i(t),j(t)}\;\; \{{\rm 
dist}_{i(t),j(t)}(t)\} \; .
\label{eq;defined(t)}
\ee
$d$ {\em is} a continuous function from $[a,b]$ to $\mathR$. For the 
continuity of worldlines, assumption b), implies that---with $i,j$ now 
labelling persisting particles!---each of the $n(n-1)/2$ functions ${\rm 
dist}_{ij}(t)$ is a continuous function of $t$, so that $d$ is also 
continuous.\footnote{If we strengthened assumption b) to say that during 
$[a,b]$, each particle  is represented by a differentiable  timelike curve  
$\gamma_i$, then $d$ would be a piecewise differentiable function. That is, it 
would be differentiable throughout $[a,b]$, except perhaps at times when which 
pair of particles was closest of all the pairs changed from one pair to 
another.}

The idea now is that the perdurantist can use $d$ to determine the worldline 
($q$-curve) through the occupied point $<t_0,x_0,y_0,z_0> \in T$. I shall 
first present the intuitive idea, then discuss how it faces a problem, and 
finally show how the corollaries to the Heine-Borel theorem solve the problem.

The intuitive idea is that since at time $t_0$, the distance of any other 
particle from the one at $<x_0,y_0,z_0>$ is at least $d(t_0)$, it follows that 
for a small interval $I_{t_0}$ of time around $t_0$ any particle closer than 
say $d(t_0)/2$ to $<x_0,y_0,z_0>$ must be the same particle as occupied 
$<x_0,y_0,z_0>$ at $t_0$. That is: during $I_{t_0}$, any other particle that 
is approaching $<x_0,y_0,z_0>$ is still only in the shell consisting of 
positions $<x,y,z>$ between the two concentric spheres around $<x_0,y_0,z_0>$:
\be
\frac{d(t)}{2} < \;\; \parallel <x_0,y_0,z_0> - <x,y,z> \parallel \;\; \equiv
\surd \left((x_0 - x)^2 + (y_0 - y)^2 + (z_0 - z)^2 \right)
 < d(t)
\label{eq;defineshell}
\ee 
One then envisages applying a similar argument at other times $t' \in 
I_{t_0}$.

 The problem with this idea, as stated, is that $I_{t_0}$ may be very small, 
as a result of a high-speed particle that will ``soon invade'' the ``inner 
sphere'' i.e. the sphere consisting of positions $<x,y,z>$ with ${\rm 
dist}(<x_0,y_0,z_0>, <x,y,z>) < d(t)/2$. More precisely, the problem is as 
follows. Unless we invoke an upper bound on particles' velocity, there is a 
risk that, as we apply the argument successively, first at $t_0$ to define 
$I_{t_0} \ni t_0$ in the way indicated, then at $t' \in I_{t_0}$ to define 
$I_{t'}$, then at $t'' \in I_{t'}$ to define $I_{t''}$ etc., the size of the 
intervals $I_{t^{(k)}}$ that we define may shrink as a result of there being 
at the times $t,t'', \dots, t^{(k)},\dots$ particles of successively higher 
velocity. (The threatened high-speed invader need not be the same particle at 
the different times.)  As a result, there is a risk that the intervals 
$I_{t^{(k)}}$ do not cover all of $[a,b]$: so that we do not determine the 
$q$-curve through $<t_0,x_0,y_0,z_0>$ for all of $[a,b]$.

If this seems just a technical glitch which is unlikely to occur in practice, 
it is worth recalling that in classical mechanics an infinite potential well, 
for example the $\frac{1}{r}$ gravitational potential around a massive 
particle, represents an inexhaustible source of energy, which another particle  
orbiting the given one might somehow tap. Furthermore, this `somehow' is 
nowadays not mere speculation. Xia (1992) proved that Newtonian gravitational 
theory  for point-particles has solutions in which particles feed off one 
another's gravitational potentials,  accelerating ever more rapidly, so that 
in a finite time-interval, say $[a,b] \subset \mathR$, they escape to spatial 
infinity. That is, they acquire arbitrarily high speed, and the worldlines do 
{\em not} register on the final time-slice $t = b$.\footnote{Xia's solution, 
using five point-particles, was the culmination of a century-long effort to 
find such solutions (called `non-collision singularities'), or to prove they 
do not exist. A very fine popular account is Diacu and Holmes (1996, Chapter 
3).} 

Indeed, even in a relativistic setting, with the strict upper bound $c$ on 
particle velocities, the above problem of shrinking intervals still looms, 
since I have not yet secured a lower bound on particles' spatial separation. 
In more detail: at speed $c$ it takes at least a time  
\be
\tau(t_0) := \frac{{\rm distance}}{{\rm speed}} := \frac{(d(t_0)/2)}{c}
\label{eq;definetau(t)}
\ee
for an invader moving at speed $c$ to traverse the shell eq. 
\ref{eq;defineshell}, i.e. to enter the inner sphere consisting of positions 
$<x,y,z>$ with $\parallel <x_0,y_0,z_0> - <x,y,z> \parallel < d(t)/2$. 
Similarly for the reverse direction of time. So: in the time-interval $I_{t_0} 
:= [t_0 - \tau(t_0), t_0 + \tau(t_0)]$, any invader (i.e. any particle 
approaching the particle that occupies $<t_0,x_0,y_0,z_0>$)  is still at worst 
only in the shell eq. \ref{eq;defineshell}. But for all I have so far said,  
the problem above remains: since $\tau = \tau(t_0)$ depends on $t_0$, the 
intervals $I_{t_0}$ might shrink so as not to cover all of $[a,b]$.

This sort of problem is familiar in elementary real analysis---and is solved 
by using the Heine-Borel theorem, or one of its corollaries. Here, the 
relevant corollary is that a continuous function on a closed bounded interval, 
such as $\tau$ on $[a,b]$, is bounded and attains its bounds. (More generally: 
the   image, under a continuous function between metric spaces,  of a compact 
set is compact; and every compact subset of a metric space is closed and 
bounded; cf. e.g. Apostol 1974, theorem 3.38, p. 63 and theorem 4.25, p. 82.)  
In fact, since the particles do not collide during the time-interval $[a,b]$ 
(assumption c) of (1) above), $d(t)$ attains a minimum during $[a,b]$ which is 
positive: let us call it $d_{{\rm min}} > 0$. Then similarly: $\tau(t)$ 
attains its minimum during $[a,b]$, which is positive: let us call it 
$\tau_{{\rm min}} = \frac{(d_{{\rm min}}/2)}{c} > 0$.

For the {\em non}-relativistic case, there is a simple strategy which adapts 
the above argument; in particular,  using the same corollary of the 
Heine-Borel theorem, that a continuous function on a closed bounded interval 
is bounded and attains its bounds. But I shall point out that this strategy is 
philosophically questionable: but nevermind---we can use another strategy 
(using another corollary), that is not questionable. 

The simple strategy applies the above corollary to the speeds of each of the 
$n$ particles. Thus we now strengthen assumption b) by assuming that each 
function $q_i$ is continuously differentiable at all $t \in [a,b]$. Then the 
speed of each of the particles $i = 1,\dots,n$ is a continuous function of 
time, and attains its maximum, say $v_i$, during $[a,b]$. Let $V$ be the 
maximum of these: $V := {\rm max}_i \{v_i \}$. Then we can argue as for the 
relativistic case, but now making $V$ take the role of the universal speed 
$c$.

 However, this strategy is philosophically questionable. For recall:\\ \indent 
(i) that our overall motivation is an interest in epistemic criteria of 
identity; or in other words, an ``engineering'' interest in answering `was the 
same particle chosen in each of two configurations?'; and\\
\indent (ii) that the notion of velocity presupposes the notion of persistence 
(cf. comment (3) in Section \ref{sssec;making precise}).\\
\indent So it seems that only once we have solved our problem i.e. determined 
the worldlines through the various  points that are given to us as occupied by 
the configuration for time $t_0$, have we any right to information about the 
$v_i$, and so about $V$. Agreed: one could reply to this objection, saying for 
example that one might have some general upper bound for $V$, like $c$ in 
relativity theory. But we do not need to explore the {\em pros} and {\em cons} 
here. For there is in any case a better strategy: one which is not 
questionable in this way, and which does not require us to strengthen 
assumption b).

Namely, we use another corollary of the Heine-Borel theorem: that a continuous 
function on a closed bounded interval is uniformly continuous on that 
interval. (More generally, a continuous function between two metric spaces is 
uniformly continuous on a compact subset of its domain: Apostol 1974, theorem 
4.47, p.91.) We apply this, not to speeds (which, with just assumption b),  
might not always exist), but to each of the particles' continuous spatial 
trajectories, i.e. the functions $q_i: [a,b] \rightarrow \mathR^3$. That is, 
for each $i = 1,\dots,n$, there is a function $\dd^i: \varepsilon \in \mathR^+ 
\mapsto \dd^i(\varepsilon) \in \mathR^+$ such that 
\be
\forall \; \varepsilon > 0, \forall \; t,t' \in [a,b]: \;\;
\mid t' - t \mid < \dd^i(\varepsilon) \;\; \Rightarrow \;\;
 \parallel q_i(t') - q_i(t) \parallel \;\; < \;\; \varepsilon \;\; ;
\label{eq;qiunifmlycts}
\ee  
where as in eq. \ref{eq;defineshell}, $\parallel \;\; \parallel$ denotes the 
usual Euclidean distance in $\mathR^3$. We now recall that since the particles 
do not collide during the time-interval $[a,b]$ the inter-particle minimum 
separation $d(t)$ attains a positive minimum during $[a,b]$, $d_{{\rm min}}$. 
We now choose for each $i$, $\frac{d_{{\rm min}}}{2}$ as the  value of 
$\varepsilon$, and we define
\be
\dd := \;\; {\rm min}\;\; \left\{\dd^1(\frac{d_{{\rm min}}}{2}), \; 
\dd^2(\frac{d_{{\rm min}}}{2}), \; \dots, \; \dd^n(\frac{d_{{\rm min}}}{2}) \; 
\right\} \; \; .
\label{eq;definecommondelta}
\ee
It follows that
\be
\forall \; t,t' \in [a,b]: \;\;
\mid t' - t \mid < \dd \;\; \Rightarrow \;\; \forall i = 1,\dots,n \;\;
\parallel q_i(t') - q_i(t) \parallel \;\; < \;\;
\frac{d_{{\rm min}}}{2} \;\; .
\label{eq;allqiless}
\ee
We now take $t$ in eq. \ref{eq;allqiless} to be $t_0$. So in the time-interval 
$[t_0, t_0 + \dd]$, the particle at $<x_0,y_0,z_0>$ can move at most 
$\frac{d_{{\rm min}}}{2}$. That is, it remains in the ``inner sphere'' of 
radius $\frac{d_{{\rm min}}}{2}$ around $<x_0,y_0,z_0>$. On the other hand, 
any other particle: (i) must at $t_0$ be at least $d_{{\rm min}}$ from 
$<x_0,y_0,z_0>$ (by the definition of $d_{{\rm min}}$); and (ii) can, in the 
time-interval $[t_0, t_0 + \dd]$,  move at most $\frac{d_{{\rm min}}}{2}$ from 
its location at $t_0$ (eq. \ref{eq;allqiless}). So any such particle cannot 
during $[t_0, t_0 + \dd]$ enter the inner sphere of radius $\frac{d_{{\rm 
min}}}{2}$ around $<x_0,y_0,z_0>$. So: we can be certain that any particle 
that is, during $[t_0, t_0 + \dd]$, in the inner sphere, is the same particle 
as was located at $<x_0,y_0,z_0>$.

\subsection{Future prospects}\label{sssec;discussionCush}
The discussion since Section \ref{ssec;reconstructCush} has been a first 
attempt to connect philosophers' concerns about persistence, especially  
epistemic criteria of identity for particles, with physicists' technical 
description  of motion. I will end with a brief list of projects suggested by 
that discussion.

First, some projects arise from the details above.\\
\indent (1): One could seek results with the ``opposite flavour'' than Section 
\ref{sssec;negresultCush}'s examples. That is, one could seek a result that 
for some class of suitably ``non-symmetric'' or ``generic'' pairs of 
configurations, a certain kind of body of information is sufficient for 
answering the question `was the same particle  chosen in the two 
configurations?'.\\
\indent (2): One could seek generalizations and analogues of Section 
\ref{sssec;posresultCush}'s result. As I mentioned there, the result's 
assumptions   can be generalized: but how exactly? The result also took it 
that we (or better: the robot's central processor) were ``given'' the 
instantaneous distances between all pairs of particles, as a function of $t$. 
That is, we were given the distances ${\rm dist}_{i(t)j(t)}(t)$; and therefore 
also the minimum function $d(t)$. But in fact such information is always 
inferred from other information; so a study of epistemic criteria of identity 
could try to model that inference. More generally, one could seek other 
results exploiting the general idea that particles  move slowly enough that 
two particles  could not swap which spatial neighbourhoods they are in. 

Finally, there are two other projects, further from the detail above, and 
closer to the concerns of engineering and physics.\\
\indent (3): One could seek results about being given a finite (discrete-time) 
sequence of configurations (as I first discussed in Section \ref{sssec;making 
precise}).\\
\indent (4): One could seek results about being given information about 
particles' velocities as well as their configurations. This last project 
brings us back to my denial of what I call `{\em pointillisme}', introduced in 
(iv) at the end of Section \ref{sec;perseintrCush}. Recall that this denial 
means that the perdurantist can  advocate only extended i.e. non-instantaneous 
temporal parts; and this means in effect that they can admit velocities into 
the supervenience basis for persistence, despite the usual circularity 
objection that the notion of velocity presupposes persistence (cf. (3) in 
Section \ref{sssec;making precise}).

So much by way of listing possible projects. In conclusion: I have described 
some questions and results in the borderlands between the philosophy of 
persistence and the physics of motion---and I hope that Jim Cushing, so wise 
in both physics and philosophy,  would have found them interesting.\\

{\em Acknowledgements}:---I thank audiences in Florence, Leeds, London and 
Oxford.

\section{References}
T. Apostol (1974), {\em Mathematical Analysis}, second edition, 
Addison-Wesley. \\
Y. Balashov (1999), `Relativistic objects', {\em Nous} {\bf 33} pp. 644-662.\\
J. Bell (1976), `The Measurement Theory of Everett and de Broglie's
Pilot Wave', in {\em Quantum Mechanics, Determinism, Causality and
Particles}, ed. M. Flato et al. Dordrecht: Reidel, pp. 11-17.
Reprinted in Bell (1987); page references to reprint.\\
J. Bell (1981), `Quantum Mechanics for Cosmologists', in {\em Quantum
Gravity II}, eds. C. Isham, R.Penrose and D. Sciama, Oxford: Clarendon
Press, pp. 611-637.  Reprinted in Bell (1987); page references to
reprint.\\
J. Bell (1987), {\em Speakable and Unspeakable in Quantum Mechanics},
Cambridge: Cambridge University Press.\\
J. Butterfield (1985), `Spatial and temporal parts', {\em the Philosophical 
Quarterly} {\bf 35}, pp. 32-44; reprinted in H. Noonan ed. (1993) {\em 
Identity}, Dartmouth. \\
J. Butterfield (2002), `The End of Time?', {\em British Journal for the 
Philosophy of Science} {\bf 53}, pp 289-330.\\
J. Butterfield (2004), `On the Persistence of Homogeneous Matter: the Rotating 
Discs Argument Defeated'; in preparation.\\
J. Butterfield (2004a), `Classical mechanics is not {\em pointilliste}, and 
can be perdurantist'; in preparation.\\  
J. Cushing (1994), {\em Quantum Mechanics: Historical Contingency and 
Copenhagen Hegemony}, Chicago University Press.\\
J. Cushing (1998), {\em Philosophical Concepts in Physics}, Cambridge  
University Press.\\
F. Diacu and P. Holmes (1996), {\em Celestial Encounters}, Princeton 
University Press.\\
P. Dowe (2000), {\em Physical Causation}, Cambridge University Press.\\
S. Haslanger (1994), `Humean supervenience and enduring things', {\em 
Australasian Journal of Philosophy} {\bf 72}, pp. 339-359.\\
K. Hawley (2001), {\em How Things Persist}, Oxford University Press.\\ 
C. Hitchcock (2003), `Of Humean Bondage', {\em British Journal for the 
Philosophy of Science} {\bf 54}, pp. 1-25.\\
M. Johnston (1987), `Is there a problem about persistence?', {\em Aristotelian 
Society Supplementary Volume} {\bf 61} pp. 107-135; reprinted in H. Noonan ed. 
(1993) {\em Identity}, Dartmouth.\\
S. Kripke (1976), `Is there a problem about substitutional quantification?', 
in {\em Truth and Meaning}, G. Evans and J. McDowell eds., Oxford University 
Press; reprinted 1999.\\
D. Lewis (1986), {\em On the Plurality of Worlds}, Oxford: Blackwell.\\
D. Lewis (1988), `Rearrangement of Particles: Reply to Lowe', {\em Analysis} 
{\bf 48} pp. 65-72; reprinted in H. Noonan ed. (1993) {\em Identity}, 
Dartmouth.\\
D Lewis (1999), `Zimmerman and the Spinning sphere', {\em Australasian Journal 
of Philosophy} {\bf 77}, pp. 209-212.\\
D.H. Mellor (1998), {\em Real Time II}, Routledge.\\
M. Rea (1998), `Temporal parts unmotivated', {\em Philosophical Review} {\bf 
107}, pp. 225-260.\\
D. Robinson (1982), `Re-identifying matter', {\em The Philosophical Review} 
{\bf 91}, pp. 317-341.\\
T. Sider (2001), {\em Four-Dimensionalism}, Oxford University Press.\\
Z. Xia (1992), 'The Existence of Noncollision Singularities in Newtonian 
Systems', {\em Annals of Mathematics} {\bf 135} pp. 411-68.\\
D.Zimmermann (1997), `Immanent causation', {\em Philosophical Perspectives} 
{\bf 11}. pp. 433-471. \\

\end{document}